\documentclass[12pt]{article}
\usepackage{amstex}
\usepackage{epsfig}

\usepackage{amstex}

\def\th{\theta}

\newcommand{\beq}{\begin{equation}}
\newcommand{\eeq}{\end{equation}}
\newcommand{\bea}{\begin{eqnarray*}}
\newcommand{\eea}{\end{eqnarray*}}
\newcommand{\beqa}{\begin{eqnarray}}
\newcommand{\eeqa}{\end{eqnarray}}
\newcommand{\vph}{\boldsymbol{\varphi}}

\newcommand{\vaa}{\boldsymbol{a}}
\newcommand{\vab}{\boldsymbol{b}}

\newcommand{\vet}{\boldsymbol{\eta}}

\newcommand{\bfe}{{\bf e}}

\newcommand{\vac}{{j{\tilde j} }}
\newcommand{\hxi}{{\hat \xi}}
\newcommand{\hbeta}{{\hat \beta}}

\newcommand{\mbar}{{\overline m}}
\newcommand{\limi}[1]{\raisebox{-0.23cm}{~\shortstack{ $\mbox{lim
}$ \\
${\vspace{-0.2cm} _{#1}}$}}}
\newcommand{\fleche}[1]{\raisebox{-0.23cm}{~\shortstack{ $\longrightarrow
$ \\
${\vspace{-0.2cm} _{#1}}$}}}

\begin{document}
\begin{flushright}\end{flushright}

\vspace{1cm}

\centerline{\Large \bf One-point functions in integrable coupled minimal models}
\vskip 1cm
\centerline{\large P.\ Baseilhac}
\vskip 0.5cm
\centerline{\it Department of Mathematics, University of York}
\centerline{\it Heslington, York YO105DD, United Kingdom}
\vskip 1.5cm
\centerline{\bf Abstract}
We propose exact vacuum expectation values of local fields for a quantum
group restriction of the $C_2^{(1)}$ affine Toda theory which
corresponds to two coupled minimal models. The central charge of the
unperturbed models ranges from $c=1$ to $c=2$, where the perturbed
 models correspond to two magnetically coupled Ising models and
 Heisenberg spin ladders, respectively. As an application, in
the massive phase we deduce the leading term of the asymptotics of the two-point
correlation functions.
\vskip 1cm
{\small\section{Introduction}}
The vacuum expectation values (VEV)s of local fields play an important
role in quantum field theory (QFT) and in statistical mechanics
\cite{1,2}. In statistical mechanics the VEVs determine the
``generalized susceptibilities'', i.e. the linear response of a
system
to external fields. Furthermore, the VEVs provide all the information
about correlation functions in QFT defined as a perturbed conformal
field theory (CFT) that is not accessible through a direct
calculation in
conformal perturbation theory \cite{3}. A few years ago, some
important
progress was made in the calculation of such quantities in integrable
(1+1) QFT. In ref. \cite{4}, an explicit expression for the VEVs of
the exponential field in the sinh-Gordon and sine-Gordon models was
proposed.
In ref.
\cite{5} it was shown that this result can be obtained using the
``reflection amplitude'' \cite{6} of the Liouville field theory. This
method was also applied in the
so-called Bullough-Dodd model with real and imaginary coupling. It is
known that $c<1$ minimal CFT perturbed  by the operators $\Phi_{12}$
and $\Phi_{13}$ can be obtained by a quantum group (QG) restriction
of the sine-Gordon \cite{7} and imaginary Bullough-Dodd model
\cite{8} with
special values of the coupling. The VEVs of primary fields were then
calculated.
The same method was applied later to integrable QFTs involving more
than
one field. For instance, the VEV for a two-parameter family of
integrable QFTs \cite{9} gave rise to the VEV of local operators in
parafermionic
sine-Gordon models and in integrable perturbed $SU(2)$ coset CFT
\cite{9bis}. The VEV of simply-laced affine Toda field
theories (ATFT)s is known for a long time \cite{norm} and
 the case of non-simply laced dual pairs was recently studied in
\cite{tba,papier2} for which a general expression
for the VEVs was derived.

Such perturbed CFTs have recently attracted much
attention in condensed matter physics, such as in the context of
point
contacts in the fractional quantum Hall effect and impurities in
quantum wires \cite{12,13}. In such cases the property of
integrability
has provide a non-perturbative answer for experimentally important
strongly interacting solid state physics problems \cite{13,14}.
Particulary, on-shell results were obtained using exact relativistic
scattering and related form factor techniques \cite{15,16}.

In this
paper, we are interested in exact off-shell results for two coupled
conformal
field theories\,\footnote{In literature, the first example of such integrable
coupled models was
studied in \cite{zN}.} for which the inter-layer coupling preserves
integrability. The on-shell dynamics of these models were studied in
\cite{zN,vays,muss}.

Let us briefly recall the ideas of \cite{zN,vays,muss}. We consider two
planar systems  corresponding to two coupled minimal
models ${\cal{M}}_{p/p'}$ which interact through a relevant operator
which
leads to an integrable theory. The resulting action can be written :
\beqa
{\cal{A}} = {\cal{M}}_{p/p'} + {\cal{M}}_{p/p'} + {\lambda} \int d^2x
\Phi^{(1)}_{12}\Phi^{(2)}_{12}\label{action}
\eeqa
or 
\beqa
{\tilde{\cal{A}}} = {\cal{M}}_{p/p'} + {\cal{M}}_{p/p'} + {\hat\lambda} \int d^2x
\Phi^{(1)}_{21}\Phi^{(2)}_{21},\label{actiontilde}
\eeqa
where  we denote respectively $\Phi^{(1)}_{12} (\Phi^{(1)}_{21})$
 and $\Phi^{(2)}_{12} (\Phi^{(2)}_{21})$
as two
specific primary operators of each unperturbed minimal models and
where the
parameters \
$\lambda$ and $\hat\lambda$ \ characterize the strength of the
 interaction. Here, we will be interested in exact one-point
functions in such system. 

In section 2 we introduce the notations and those known results which
are useful for
our purpose. In section 3 using the exact
result for the VEV of the $C_2^{(1)}$ ATFT derived in \cite{papier2},
we deduce the exact VEV \ $<\Phi^{(1)}_{rs}(x)\Phi^{(2)}_{r's'}(x)>$
 for any values of $(r,s),(r',s')$ in the model with action
(\ref{action}). To do it, we relate the parameter $\lambda$ in
(\ref{action}) to the masses of the particles and we perform
the QG restriction of the $C_2^{(1)}$ ATFT with
imaginary coupling which leads to the model (\ref{action}). For
$(r,s)=(r',s')=(1,2)$ this VEV can be calculated exactly as well as the
bulk free
energy. The specific case ${\cal M}_{3/4}+{\cal M}_{3/4}$ coupled by
$\Phi^{(1)}_{12}\Phi^{(2)}_{12}$ is considered in details. It corresponds to two layer Ising models
coupled by their magnetization operator $\sigma^{(1)}\sigma^{(2)}$. The
previous approach is also extended to the model described by action 
(\ref{actiontilde}). In section 4, we extract some limited information
about the asymptotics of two-point correlation functions between any
 pairs of primary operators which belong to the same or
 different unperturbed models. More precisely, we will distinguish four different cases :
\beqa
 (a) &:& <\Phi^{(1)}_{rs}(x)\Phi^{(2)}_{r's'}(y)> \ \ \ \mbox{for} \
\
|x-y|\rightarrow 0,\nonumber \\
 (b) &:& <\Phi^{(1)}_{rs}(x)\Phi^{(2)}_{r's'}(y)> \ \ \ \mbox{for} \
\
|x-y|\rightarrow \infty, \nonumber\\
 (c) &:& <\Phi^{(i)}_{rs}(x)\Phi^{(i)}_{r's'}(y)> \ \ \ \mbox{for} \
\
|x-y|\rightarrow 0\ \ \ \mbox{and} \ \ i\in\{1,2\}, \nonumber\\
 (d) &:& <\Phi^{(i)}_{rs}(x)\Phi^{(i)}_{r's'}(y)> \ \ \ \mbox{for} \
\
|x-y|\rightarrow \infty\ \ \ \mbox{and} \ \ i\in\{1,2\}\nonumber
\eeqa
which are depicted in figure 1.
We finally give some numerical results for various examples of coupled
minimal models such as two energy-energy coupled tricritical Ising, two coupled $A_5$-RSOS
models and two energy-energy coupled 3-state Potts model.
Perspective and conclusions follow in this final section.

\vspace{5mm}

\centerline{\epsfig{file=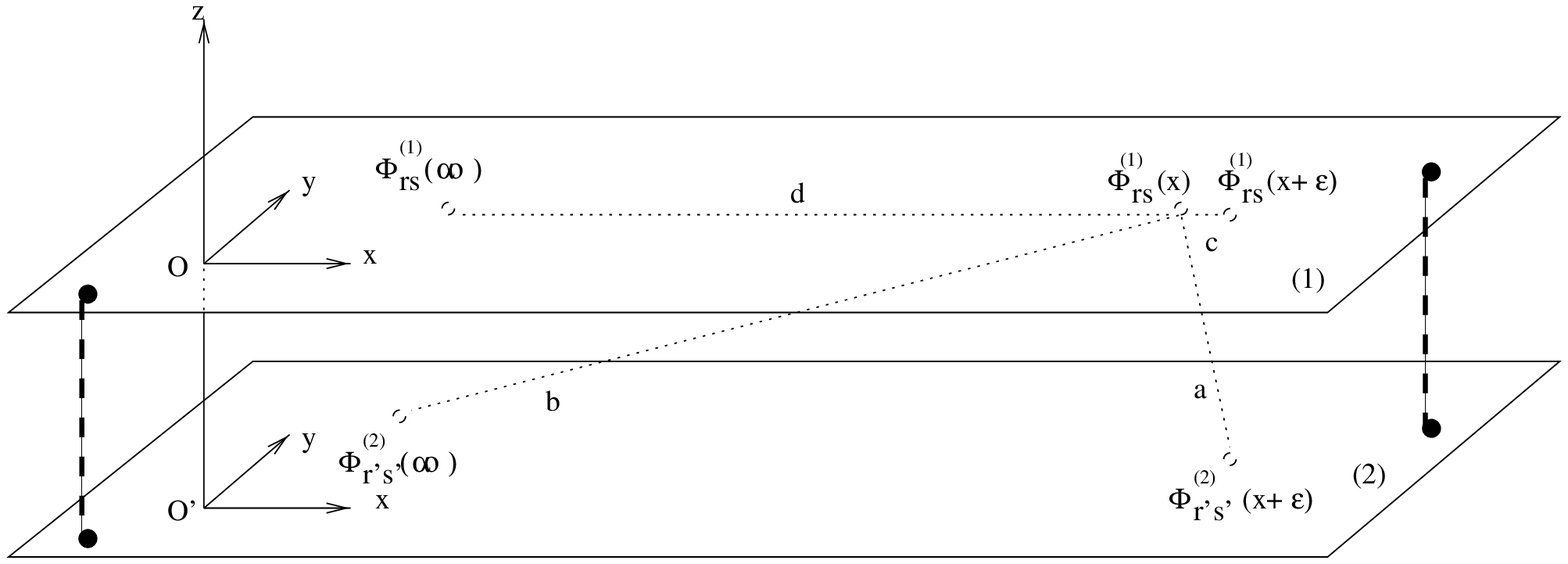,height=55mm,width=155mm}}
\vspace{5mm}
\begin{center}
{\small \underline{Figure 1}  - Two coupled two dimensional models.
Short
distance}\\
{\small \ \ \ \ \ results are obtained by taking the limit \
$\epsilon\rightarrow 0$.}
\end{center}

\vspace{0.1cm}

{\small\section{Coupled minimal models as restricted $C_2^{(1)}$ ATFT}}
The ATFT with real coupling $b$ associated with the affine Lie
algebra
 $C_2^{(1)}$ is described by the action in the Euclidean space :
\beqa
{\cal A}=\int d^2x\big[\frac{1}{8\pi}(\partial_\mu\vph)^2 +
 \mu'e^{-2b\varphi_1} + \mu'e^{2b\varphi_2} + \mu
e^{b(\varphi_1-\varphi_2)}
\big]\label{action2}
\eeqa
where we choosed the convention that the length squared of the long
roots is four. As the different vertex operators
do not
renormalize in the same way, we introduced two scale parameters $\mu$
and $\mu'$. The fields in eq. (\ref{action2}) are normalized such
that:
\beqa
<\varphi_i(x)\varphi_j(y)>=-\delta_{ij}\ln|x-y|^2.\label{prop}
\eeqa
This model possess two fundamental particles with mass $M_a$ which depends
on one parameter $\mbar$
:
\vspace{-0.3cm}
\beqa
M_a =2\mbar\sin(\frac{\pi a}{H}) \ \ \ \ \mbox{for} \ \ a=1,2  \label{fund}
\eeqa
where we introduced the ``deformed'' Coxeter\,\footnote{Differently to the
simply laced case for which the mass ratios take the classical
 values, the mass ratios for non-simply laced case get quantum
corrections.} number \cite{delius} $H=h(1-B)+h^\vee B$
with $B=\frac{b^2}{1+b^2}$ and $h=4$, $h^\vee=6$ are respectively the
 Coxeter and dual Coxeter numbers. The exact relation between $\mbar$,
$\mu$ and $\mu'$ was found in \cite{tba} and is given by :
\beqa
\big(-\pi \mu \gamma(1+b^2)\big) \big(-\pi \mu'
\gamma(1+2b^2)\big)
= 2^{-\frac{4}{1-B}}\big(\mbar\frac{\Gamma((1-B)/H)\Gamma(1+B/H)}
{\Gamma(1/H)}\big)^{\frac{H}{1-B}}\label{mass-mu}.
\eeqa

In ATFT approach to perturbed CFT, one usually identifies the
perturbation with the affine extension of the Lie algebra $\cal G$.
Instead, the perturbation will be associated here with the standard (length
2)
root of $C_2^{(1)}$. Removing the last term in the action (\ref{action2})
leaves a $D_2=SO(4)=SU(2)\otimes SU(2)$ model, i.e. two decoupled Liouville
models.
To associate the two first terms of the $C_2^{(1)}$ Toda potential to
two decoupled conformal field theories, we first introduce for each one a
background charge at infinity. Then, the total stress-energy tensor
 $T(z)$ is written $T(z)=T^{(1)}(z)+T^{(2)}(z)$ where :
\beqa
T^{(i)}(z)=-\frac{1}{2}(\partial\varphi_i)^2 +
Q_i\partial^2\varphi_i\ \ \
\mbox{for} \ \ i\in \{1,2\}
\eeqa
ensures the local conformal invariance of the $D_2$ model
for the
choice
$Q_2=-Q_1=b+1/2b$.
With our conventions\,\footnote{Here, the length of the longest
root is chosen to be 4.}, the exponential fields
\beqa
e^{a_i\varphi_i(x)}\ \ \ \ \mbox{for} \ \ \ \ i\in
\{1,2\}\label{expo}
\eeqa
are spinless conformal primary fields of each Liouville model with
conformal dimensions
\beqa
\Delta(e^{a_i\varphi_i(x)})=-\frac{a_i^2}{2}+a_iQ_i \ \ \ \
\mbox{and}\
\ \ \ i\in \{1,2\}.
\eeqa
In particular, the exponential fields $e^{-2b\varphi_1}$ and
 $e^{2b\varphi_2}$ have conformal dimensions 1. As is well known, the
``minimal model''
${\cal M}_{p/p'}$ with central charge $c=1-6\frac{(p-p')^2}{pp'}$ can
be
obtained from the Liouville case. Consequently, the $D_2$
 CFT can be identified with two decoupled minimal models by the
substitution :
\beqa
b\rightarrow i\beta,\ \ \ \ \ \ \mu\rightarrow -\mu, \ \ \ \ \ \
\mu'\rightarrow -\mu',\label{sub}
\eeqa
and the choice :
\beqa
\beta^2=\beta^2_+=p/2p' \ \ \ \ \mbox{or}\ \ \ \
\beta^2=\beta^2_-=p'/2p \ \ \ \ \mbox{with}\ \ \ \ p<p'.\label{restrict}
\eeqa
Similarly, the primary operators of each minimal model
 ${\cal M}_{p/p'}$ are obtained through the substitution $a_j \rightarrow
i\eta_j$
 for $j=1,2$ in (\ref{expo}).
With these substitutions, the conformal dimension of the perturbing
 operator becomes :
\beqa
\Delta_{pert}=\Delta(e^{i\beta(\varphi_1-\varphi_2)})=3\beta^2-1.
\eeqa
As long as we consider a relevant perturbation,
we are restricted to choose $\beta^2<2/3$. In the following we will
consider only the cases for which this condition is satisfied,
in particular $\beta^2=\beta^2_+$.

We define respectively $\{\Phi^{(1)}_{rs}\}$ and
$\{\Phi^{(2)}_{r's'}\}$ as the
two sets of primary fields with conformal dimensions :
\beqa
\Delta_{rs}=\frac{(p'r-ps)^2-(p-p')^2}{4pp'} \ \ \ \ \mbox{for}\ \
\ 1\leq r<p,\ \ 1\leq s < p' \ \ \mbox{and}\ \ \ p<p'.\label{dim}
\eeqa
They are simply related to the vertex operators of each minimal model
through the relation:
\beqa
\Phi_{rs}^{(i)}(x)=N^{(i)-1}_{rs}\exp(i\eta_i^{rs}\varphi_i(x))\ \ \
\ \ \
\mbox{with}\ \ \ \ \ \ \ \
\eta_1^{rs}=-\eta_2^{rs}=\frac{(1-r)}{2\beta}-(1-s)\beta,\label{primaire}
\eeqa
where we have introduced the normalization factors $N^{(i)}_{rs}$ for
each model. These numerical factors depend on the
normalization
of the primary fields. Here, they are chosen in such a way that the
primary
fields satisfy the conformal normalization condition :
\beqa
<\Phi^{(i)}_{rs}(x)\Phi^{(i)}_{rs}(y)>_{CFT} \ =\
\frac{1}{|x-y|^{4\Delta_{rs}}}
 \ \ \ \mbox{for} \ \ \ \ i\in \{1,2\}.
\eeqa
For further convenience, we write these
coefficients $N^{(i)}_{rs}=N^{(i)}(\eta_i^{rs})$ where :
\beqa
N^{(1)}(\eta)=\big[-\pi\mu'\gamma(-2\beta^2)\big]^{\frac{\eta}{2\beta}}
\big[\frac{\Gamma(2\beta^2+2\eta\beta)\Gamma(1/2\beta^2-\eta/\beta)
\Gamma(2-2\beta^2)\Gamma(2-1/2\beta^2)}{\Gamma(2-2\beta^2-2\eta\beta)
\Gamma(2-1/2\beta^2+\eta/\beta)\Gamma(2\beta^2)\Gamma(1/2\beta^2)}\Big]^{\frac{
1}{2}}
\nonumber
\eeqa
and $N^{(2)}(\eta)=N^{(1)}(-\eta)$.

For imaginary values of the coupling $b=i\beta$, the
$C_2^{(1)}$ ATFT possesses complex soliton solutions which interpolate
between the
degenerate vacua. This QFT possesses the QG symmetry associated to
$U_q(D_3^{(2)})$ - as we will recall
in the next section. In
 \cite{MKay} the $S$-matrix of the $B_2^{(1)}$ ATFT was constructed using
the
$U_q(A_3^{(2)})$ QG symmetry of this QFT. In particular, using the
 bootstrap procedure, the authors deduced the breather-breather
$S$-matrix. It was also shown that a breather-particle
 identification holds by comparing the $S$-matrix elements of the
lowest-breathers (breathers with lowest mass) with the $S$-matrix
elements  for
the quantum particles in real ATFT. Following the conventions of
\cite{tba}, the particle spectrum in real $B_2^{(1)}$ ATFT is given by
 \ $M_1=2\mbar\sin(\pi/H)$ \ and \ $M_2=\mbar$. Using the results of
\cite{MKay} we find the identification :
\beqa
\mbar = 2M\sin(\frac{\pi\xi}{4-2\xi})\ \ \ \ \mbox{with}\ \ \ \
 \ \xi=\frac{\beta^2}{1-\beta^2}\label{rel}
\eeqa
where we denote $M$ as the mass of the lowest kink. Eq. (\ref{rel})
 holds for our case due to the identification $B_2^{(1)}\equiv
C_2^{(1)}$ and $A_3^{(2)}\equiv D_3^{(2)}$.\\

{\small\section{Expectation values in coupled minimal models}}
In ref. \cite{tba,papier2} we derived the exact VEV \ \
$G(\vaa)=<\exp(\vaa.\vph)(x)>$\ \ for all non-simply laced ATFTs
using the so-called ``reflection relations'' which relate different fields
with
the same quantum numbers. We refer the reader to these papers for more
details. Although the model (\ref{action}) is very different from the
$C_2^{(1)}$
ATFT (\ref{action2}) in its physical content (the model
(\ref{action}) contains solitons and excited solitons), there are good
reasons to believe that the expectation values obtained in the real
coupling case provide also the expectation values for imaginary coupling.
The calculation of
the VEVs in both cases ($b$ real or imaginary) within the standard
perturbation theory agree through the identification $b=i\beta$.
Following the analysis done for the Bullough-Dodd model \cite{5}, it
 is then straightforward to obtain the VEV of
primary operators which belong to different minimal models. With the
substitutions (\ref{sub}) one gets ${\cal G}(\vet)=G(i\vet)$ :
\beqa
{\cal G}(\vet)&=&
\Big[\pi^2\mu\mu'\gamma(\frac{1}{1+\xi})\gamma(\frac{1-\xi}{1+\xi})\Big]^{\frac
{(1+\xi)}{2-\xi}(\Delta_1+\Delta_2)}
\Big[\pi\mu'\gamma(\frac{1-\xi}{1+\xi})\Big]^{\frac{(1+\xi)}{2\xi}\beta(\eta_1-
\eta_2)}
\Big[\frac{1}{1+\xi}\Big]^{-\vet^2}\nonumber\\
&& \times \ \exp\Big[\int_0^{\infty} \frac{dt}{t}
\Big(\frac{\chi(\vet,t)}{\sinh((1+\xi)t)\sinh(2t\xi)\sinh((4-2\xi)t)}
- \vet^2 e^{-2t}\Big) \Big]\label{Geta}
\eeqa
with
\beqa
\chi(\vet,t)&=&2\big[\sinh((\eta_1-\eta_2)\beta(1+\xi)t)\sinh(((\eta_1-\eta_2)\beta
(1+\xi)+2\xi-2)t)\nonumber\\
&&\ \ \
+\sinh^2((\eta_1+\eta_2)\beta(1+\xi)t)\Big]\sinh(t)\cosh(t\xi)\nonumber
\\
&&+\Big[\sinh(2\eta_2\beta(1+\xi)t)\sinh((2\eta_2\beta(1+\xi)-2)t)\nonumber\\
&&\ \ \ +
\sinh(2\eta_1\beta(1+\xi)t)\sinh((2\eta_1\beta(1+\xi)+2)t)\Big]\sinh(
(1-\xi)t).\nonumber
\eeqa
As usual we denote
$\gamma(x)=\Gamma(x)/\Gamma(1-x)$. The integral in (\ref{Geta}) is
convergent if :
\beqa
-\frac{2}{(1+\xi)}<(\eta_1+\eta_2)\beta<\frac{2}{(1+\xi)}\ \  \ \ \mbox{and}
\
\ \
\ -1<(\eta_1-\eta_2)\beta<\frac{3-\xi}{(1+\xi)},\label{cond1}
\eeqa
and is defined via analytic continuation outside this domain.
In (\ref{Geta}) we defined $\Delta_1$ and $\Delta_2$, the conformal
dimensions in the imaginary coupling case by
$\Delta_1=\Delta(e^{i\eta_1\varphi_1
})=\frac{\eta_1^2}{2}
+\frac{\eta_1\beta}{2}(\frac{\xi-1}{\xi})$ and similarly for
$\Delta_2$ with
the change
$\eta_1\rightarrow-\eta_2$.

If we want to express the final result for the VEV in terms of the parameter
$\lambda$ in the action (\ref{action}), we need the exact
relation
between $\lambda$ and the parameters $\mu,\mu'$ in the $C_2^{(1)}$
ATFT
with imaginary coupling. We obtain :
\beqa
\lambda=\frac{\pi\mu\mu'}{(4\beta^2-1)^2}\gamma(4\beta^2)
\gamma^2(1-2\beta^2),\label{lambda}
\eeqa
which corresponds to $\lambda=-\mu N^{(1)}(\beta)N^{(2)}(-\beta)$.

Like any other ATFT, the model (\ref{action2}) for imaginary values of
the coupling has non-local conserved charges
$\{Q_k,{\overline Q}_k\}$ for $k=0,1,2$ generated respectively by the
purely chiral and anti-chiral components :
\beqa
J_{\bfe_k^\vee}(z)=e^{-\frac{i}{\beta}\bfe_k^\vee.\varphi(z)}\ \ \ \ \ \ \
\mbox{and}\ \ \ \ \ \ \ \ \ \ {\overline
J}_{-\bfe_k^\vee}({\overline
z})=e^{-\frac{i}{\beta}\bfe_k^\vee.{\overline
\varphi}({\overline z})},
\eeqa
where the fundamental vector field $\vph(z,{\overline
z})=\varphi(z)+{\overline \varphi}({\overline
z})$ and $\{\bfe_k^\vee\}$, $k=0,1,2$ is the set of dual simple roots of
the non-simply laced affine Lie algebra $C_2^{(1)}$. We also define the
topological charge :
\beqa
H_k=-\frac{\beta}{2i\pi}\int d^2x\ \bfe_k^\vee.\partial_x\vph(x,t).
\eeqa
Using the equal-time braiding relations for all $x,y$ :
\beqa
J_{\bfe_k^\vee}(x,t) {\overline J}_{-\bfe_l^\vee}(y,t) = q_l^{-a_{kl}}
{\overline
J}_{-\bfe_l^\vee}(y,t)J_{\bfe_k^\vee}(x,t)\ \ \ \ \ \mbox{with}
\ \ \ q_l=e^{-\frac{2i\pi}{|\bfe_k|^2\beta^2}},
\eeqa
where $a_{kl}=\frac{2\bfe_k.\bfe_l}{|\bfe_k|^2}$ denotes the extended Cartan
matrix
of $C_2^{(1)}$, one can show that the charges $\{Q_k,{\overline
Q}_k,H_k\}$ for $k=0,1,2$ satisfy the quantum universal enveloping
algebra $U_q(D_3^{(2)})$ with deformation parameter $q\equiv
q_0=e^{-\frac{i\pi}{2\beta^2}}$. If
we express these generators in terms of the standard Chevalley basis
$\{E_k^+,E_k^-,H_k\}$ by :
\beqa
Q_k\equiv E_k^+ q^{H_k}\ \ \ \ \ \mbox{and}\ \ \ \ \ \ {\overline Q}_k\equiv
E_k^-
q^{H_k} \ \ \ \ \mbox{for} \  k=0,1,2,
\eeqa
then we have :
\beqa
&&[H_k,H_l]=0,\\
&&[H_k,E^\pm_l]=\pm a_{l,k}E^\pm_l, \nonumber\\
&&[E^+_k,E^-_l]=\delta_{kl}\frac{q^{2H_l}-q^{-2H_l}}{q_l-q_l^{-1}}\label{alg}.\nonumber
\eeqa

The $U_q(D_3^{(2)})$ has two subalgebras $U_q(D_2)$ as well as
$U_{q^2}(C_2)$ \cite{MKay} where the subalgebra $U_q(D_2)$ is generated
by $\{Q_0,{\overline Q}_0, Q_2,{\overline Q}_2,H_0,H_2\}$. As was found
in \cite{MKay} the $S$-matrix in the unrestricted form acts as an
intertwinner on the modules of the $U_q(D_2)$ representations :
\beqa
S_{a,b}(\theta)\ \ : \ \ \ {\cal V}_{\rho_a} \otimes  {\cal V}_{\rho_b}
\longrightarrow
{\cal V}_{\rho_b} \otimes  {\cal V}_{\rho_a}
\eeqa
where ${\cal V}_\rho$ is the module with highest weight $\rho$. The
scattering of two solitons of species $a$ and $b$ is then described by
$S_{a,b}(\theta_{12})$ with $\theta_{12}$ being their relative rapidity.
There are two such fundamental multiplets denoted $\{4\}$ and $\{6\}$ in
ref. \cite{muss}. In addition, there are also scalar bound states and
excited solitons depending on the values of $(p,p')$ chosen.

To understand the restricted $C_2^{(1)}$ (denoted $RC_2^{(1)}$ below), we
use the general framework of superselection sectors (for details see
\cite{fro,fre,fel}).
The model
(\ref{action}) is a perturbation of the two minimal models by the
operator $\Phi^{(1)}_{12}\Phi^{(2)}_{12}$. Each minimal model contains
a finite number of primary fields (\ref{dim},\ref{primaire}). Using the
superselection criterion for the present case\,\footnote{We only
consider the holomorphic part of the primary operators but keep the same
notation.}:
\beqa
\big(\Phi^{(1)}_{2j+1\ 1}(z)\Phi^{(2)}_{2{\tilde j}+1\
1}(z)\big)\big(\Phi^{(1)}_{12}(w)\Phi^{(2)}_{12}(w)\big)
\sim \frac{1}{(z-w)^{l}}\Phi^{(1)}_{2j+1\ 2}(w)\Phi^{(2)}_{2{\tilde j}+1\
2}(w),\ \ \ \ l\in{\mathbb Z}
\eeqa
we find\ \ $j+{\tilde j}\in{\mathbb Z}$\ \ where $\vac$ denote the
representations of $U_q(D_2)$ with $j$ the spin-$j$ representation of
$SU(2)$ with dimension $2j+1$ (and similarly for ${\tilde j}$).
If $q$ is a root of unity i.e if eq. (\ref{restrict}) is satisfied then
using the Coulomb gas representation condition (\ref{dim}) $1\leq 2j+1\leq
p-1$ and
$1\leq 2{\tilde j}+1\leq p-1$ one has :
\beqa
0\leq j\leq p/2-1\ \ \ \ \ \ \mbox{and}\ \ \ \ \ \ 0\leq {\tilde j}\leq
p/2-1.
\eeqa
The superselection sectors ${\cal H}_\vac^{RC_2^{(1)}}$ of the $RC_2^{(1)}$
model (\ref{action}) are thus :
\beqa
{\cal H}^{RC_2^{(1)}}=\sum_{(j,{\tilde j})\in\{0,1/2,...,p/2-1\}}{\cal
H}_{\vac}^{RC_2^{(1)}}\ \ \ \ \ \mbox{with}\ \ \ j+{\tilde j}\in{\mathbb Z}.
\eeqa
As shown in \cite{vays,muss} for the unitary series ($p'=p+1$), after the
quantum group retriction (\ref{restrict}) for $p>3$ the fundamental
solitons in the $\{4\}$ and $\{6\}$ representation of $U_q(D_2)$ become
the RSOS kink $K^{\vac}_{j_2{\tilde j}_2\  j_1{\tilde j}_1}$. These kinks
interpolate between different vacua $|0_{j_1{\tilde j}_1}>$ and
$|0_{j_2{\tilde
j}_2}>$ which are connected using the $U_q(sl_2)$
fusion ring at $q=-e^{-i\frac{\pi}{p}}$:
\beqa
j_1 \times j = \sum_{j_2=|j_1-j|}^{min(j_1+j,p-2-j_1-j)} j_2
\eeqa
and similarly for ${\tilde j}$. However, for $p=3$, $(j,{\tilde
j})\in\{0,1/2\}$ and then the $\{6\}$ is projected out of the spectrum,
 leaving only the $\{4\}$. We refer the reader to \cite{vays,muss}
for more details.

From the previous remarks and the identification
$D_2=SO(4)=SU(2)\otimes SU(2)$, by analogy with \cite{5} the primary fields
$\Phi^{(1)}_{1s}$ and $\Phi^{(2)}_{1s'}$ commute with the generators
in  (\ref{alg}) (for $k=0,2$) of the subalgebra $U_q(D_2)\subset
U_q(D_3^{(2)})$.
If one interpret the fields $\Phi^{(1)}_{2j+1\ 1}(z)\Phi^{(2)}_{2{\tilde
j}+1\
1}(z)$ as the highest component fields in the multiplet, it can be shown
that the primary operators
 $\Phi^{(1)}_{rs}$ and $\Phi^{(2)}_{r's'}$ are not invariant with
respect to  $U_q(D_2)$. Together with some non-local fields they form
finite-dimensional representation of this algebra. Consequently, the VEV in
(\ref{action}) should take into account the factor \ $d_{rs,r's'}^{\vac}$\
coming from the QG restriction of the QFT (\ref{action2}). Following the
conjecture of \cite{5} it takes the form :
\beqa
d_{rs,r's'}^{\vac}=\frac{\sin(\frac{\pi(2j+1)}{p}|p'r-ps|)}{\sin(\frac{\pi(2j+1)}{p}(p'-p))}
\frac{\sin(\frac{\pi(2{\tilde j}+1)}{p}|p'r'-ps'|)}{\sin(\frac{\pi(2{\tilde
j}+1)}{p}(p'-p))}.\label{d}
\eeqa
Using the notations introduced in the previous section, and eqs.
(\ref{primaire}), (\ref{Geta}),
the outcome for the VEV between different primary operators is :
\beqa
<0_\vac|\Phi^{(1)}_{rs}(x)\Phi^{(2)}_{r's'}(x)|0_\vac> &=&
d_{rs,r's'}^{\vac}
\Big[\frac{-\pi\lambda\gamma(\frac{1}{1+\xi})(1+\xi)^{\frac{4-2\xi}{1+\xi}}}
{\gamma(\frac{3\xi-1}{1+\xi})\gamma(\frac{1-\xi}{1+\xi})}
 \Big]^{\frac{(1+\xi)}{2-\xi}(\Delta_{rs}+\Delta_{r's'})}
 \nonumber
\\
&&\times \exp{\cal Q}_{12}((1+\xi)r-2\xi s,(1+\xi)r'-2\xi s').\label{VEV1}
\eeqa
The function  \ ${\cal Q}_{12}(\th,\th')$ \ for\  $|\th\pm\th'|<4\xi$ and \ $\xi>
\frac{1}{3}$ \ is given by the integral :
\beqa
{\cal Q}_{12}(\th,\th')= \int_0^{\infty} \frac{dt}{t}
\Big(\frac{\Psi_{12}(\th,\th',t)}{\sinh((1+\xi)t)\sinh(2t\xi)\sinh((4-2\xi)t)}
-\frac{\th^2+{\th'}^2 - 2(1-\xi)^2}{4\xi(\xi+1)}e^{-2t}
\Big)\nonumber
\eeqa
with
\beqa
\Psi_{12}(\th,\th',t)&=&\Big[\cosh((\th+\th')t)\cosh((\th-\th')t)\nonumber\\
&&\ \ \ \ \ -\
\cosh((2-2\xi)t)\Big]\sinh((1-\xi)t)\cosh((4-2\xi)t)\nonumber\\
&&+\Big[\cosh((\th+\th')t)+\cosh((\th-\th')t)-\cosh((2-2\xi)t)-1\Big]
\sinh(t)\cosh(t\xi).\nonumber
\eeqa
and defined by analytic continuation outside this domain.
Notice that eq. (\ref{VEV1}) satisfies:
\beqa
<0_\vac|\Phi^{(1)}_{rs}(x)\Phi^{(2)}_{r's'}(x)|0_\vac>=
<0_\vac|\Phi^{(1)}_{p-r\ p'-s}(x)\Phi^{(2)}_{p-r'\ p'-s'}(x)|0_\vac>.
\eeqa

A particular case of eq. (\ref{VEV1}) is the expectation value of the
perturbing operator which can be calculated explicitly to give the
result :
\beqa
<0_{\vac}|\Phi^{(1)}_{12}(x)\Phi^{(2)}_{12}(x)|0_{\vac}> &=&
\frac{1}{\lambda}
\Big[\frac{-\pi\lambda\gamma(\frac{1}{1+\xi})}
{\gamma(\frac{3\xi-1}{1+\xi})\gamma(\frac{1-\xi}{1+\xi})}
\Big]^{\frac{1+\xi}{2-\xi}}
\frac{
2^{\frac{3\xi-2}{2-\xi}}(1+\xi)}{\pi(\xi-2)}\frac{\gamma(\frac{\xi}{4-2\xi})\gamma(\frac{1}{4-2\xi})}
{\gamma(\frac{1+\xi}{4-2\xi})}.\label{VEV2}
\eeqa
By using the exact relation between the mass parameter $\mbar$ and the
 mass of the kink $M$
 (\ref{rel}) and eqs. (\ref{mass-mu}), (\ref{lambda}) we immediately obtain the relation
between $M$ and $\lambda$ :
\beqa
M=\frac{2^{\frac{\xi}{2-\xi}} \Gamma(\frac{\xi}{4-2\xi})
\Gamma(\frac{1}{4-2\xi})}{\pi\Gamma(\frac{1+\xi}{4-2\xi})}\Big[
\frac{-\pi\lambda\gamma(\frac{1}{1+\xi})
}{\gamma(\frac{3\xi-1}{1+\xi})\gamma(\frac{1-\xi}{1+\xi})}
\Big]^{\frac{1+\xi}{4-2\xi}}.\label{M}
\eeqa
Consequently, according to eqs. (\ref{lambda}), (\ref{VEV2}), (\ref{M}) and
$\beta^2<2/3$,
the perturbed CFTs (\ref{action}) develop a massive spectrum for
(${\mathfrak
I}m(\lambda)=0$) :
\beqa
(i) \ \ \  0<\xi<1/3, \ \ \ \ \  \lambda > 0 \ \ \ \
\mbox{i.e.}\ \ \ 0 < \frac{p}{p'} <\frac{1}{2},\\
(ii) \ \ \  1/3<\xi<1, \ \ \ \ \  \lambda < 0 \ \ \ \
\mbox{i.e.}\ \ \ \frac{1}{2} < \frac{p}{p'} <1.\nonumber
\eeqa
where $\xi=\frac{p}{2p'-p}$. In particular, the condition $(ii)$ is always satisfied
for the coupled unitary minimal  models defined by (\ref{action}).

Finally, the expectation value (\ref{VEV2}) can be used
to derive the bulk free energy :
\beqa
f_{12} = -\limi{V\rightarrow
\infty}\frac{1}{V} \ln Z,
\eeqa
where $V$ is the volume of the 2D space and $Z$ is the singular part
of the partition function associated with action (\ref{action}).
Using
\beqa
\partial_\lambda f_{12}\ =\
<0_\vac|\Phi^{(1)}_{12}\Phi^{(2)}_{12}|0_\vac>,\label{free}
\eeqa
and eqs. (\ref{VEV2}), (\ref{M}), the result
for the
bulk free energy follows :
\beqa
f_{12}=-\frac{M^2\sin(\frac{\pi}{4-2\xi})}{2}\frac{\sin(\frac{\pi\xi}{4-2\xi})}{\sin
(\frac{\pi(1+\xi)}{4-2\xi})}\label{fbulk}.
\eeqa

As was suggested in \cite{vays,muss}, a particular case of the model
(\ref{action}) can be
related to the SG theory at the reflectionless point as well as the
$D_8^{(1)}$ ATFT with imaginary coupling. Let us first consider
the SG theory with action :
\beqa
{\cal A}_{SG}=\int d^2x \big[\frac{1}{16\pi}(\partial_\nu \phi)^2
+\Lambda\cos(\hbeta\phi)\big].\label{SG}
\eeqa
This model is integrable and its on-shell solutions, i.e.
spectrum of particles and $S$-matrix are well-kown \cite{Za-Za}. This
theory contains soliton, antisoliton and soliton-antisoliton bound-state
(breathers) $B_n$, $n=1,...,1/{\hxi}$ where
$\hxi=\frac{\hbeta^2}{1-\hbeta^2}$. The lightest
bound-state $B_1$ coincides with the field $\phi$ in the perturbative
treatment of the QFT (\ref{SG}). Its mass is given by
$m_1=2M_{SG}\sin(\frac{\pi\hxi}{2})$ where $M_{SG}$ denotes the soliton
mass of the SG model. For $p=3$, $p'=4$ the model (\ref{action})
corresponds to two
magnetically coupled Ising models. It is a $c=1$ CFT perturbed by the
operator
$\Phi^{(1)}_{12}\Phi^{(2)}_{12}$ with conformal dimension $1/8$. Indeed,
the VEV of this operator must be simply related to the VEV of
$<\cos(\hbeta\phi)>$ in the SG model for $\hbeta^2=1/8$, which is also a
$c=1$ CFT with
perturbing operator of conformal dimension $1/8$. For this latter
value, $\hxi=1/7$ and the SG model possesses 6 neutral excitations :
\beqa
&&m_a=2M_{SG}\sin(\frac{\pi a}{14}),\ \ \ \ \  a=1,...,6\ ;\label{specsg}\\
&&m_1<m_2<M_{SG}<m_3<...<m_6.\nonumber
\eeqa
Similarly, for $\beta^2=3/8$, $\xi=3/5$
the spectrum of the model (\ref{action}) was described in \cite{muss}
and summarize here.
 Here,
we denote the lowest mass by $M$. If we compare the spectrum of each model,
using the notations of
\cite{MKay} we have :
\beqa
m_1\equiv M,\ \ \ \ m_2 \equiv m_{B_{1}^{(1)}},\ \ \ \ M_{SG}\equiv
m_{B_1^{(2)}},\ \ \ \mbox{etc}...\label{ident}
\eeqa
Furthermore, the bulk free energy of the SG model (\ref{SG}) is well known
\cite{Dev,4} and given by :
\beqa
f_{SG}=-\frac{M^2_{SG}}{4}\tan(\frac{\pi\hxi}{2}).
\eeqa
By expressing it in terms of the lightest particle $m_1$ using
(\ref{specsg}) and the relation\\
$8\sin(\pi/14)\sin(3\pi/14)\cos(\pi/7)=1$ we find the same result as in
eq. (\ref{fbulk}) for $\beta^2=3/8$, i.e. $\xi=3/5$. Also, using
the results of \cite{4} for $\hbeta^2=1/8$, one has :
\beqa
\Lambda=-\frac{2\gamma(1/8)}{\pi}\Big[
\frac{M_{SG}{\sqrt\pi}}{2}\frac{\Gamma(4/7)}{\Gamma(1/14)}\Big]^{\frac{7}{4}}.\label{31}
\eeqa
Similarly, the VEV of the perturbing operator is given by :
\beqa
<\cos(\hbeta\phi)>_{\hbeta^2=1/8}=\frac{8}{7}\frac{\Gamma(1/7)\Gamma(6/7)}
{\Gamma^2(4/7)\Gamma^2(13/14)}\Big[\frac{\pi}{\gamma(1/8)}\Big]^{8/7}(-\frac{\Lambda}{2})^{1/7}.\label{32}
\eeqa
Comparing eqs. (\ref{31}), (\ref{32}) with eqs. (\ref{VEV2}) and
(\ref{M}) respectively for $\xi=3/5$ and
using eq. (\ref{ident}), one find the relations :
\beqa
\Lambda \equiv 2^{\frac{1}{2}}\lambda \ \ \ \ \ \mbox{and}\ \ \ \ \ \
\cos(\hbeta\phi)|_{\hbeta^2=1/8}\equiv
2^{-\frac{1}{2}}\Phi^{(1)}_{12}\Phi^{(2)}_{12}
\eeqa
in perfect agreement\,\footnote{At $\hbeta^2=1/8$, the operator can be
expressed \cite{4,McCoy} in terms of two independent Ising spin fields,
 $\sigma^{(i)}=\Phi_{12}^{(i)}$ for $i=1,2$.} with eq. (B.21) in
\cite{4}.

Let us now turn to the relation with the $D_8^{(1)}$ ATFT. The
calculation of the bulk free energy for the simply-laced $D_8^{(1)}$
ATFT gives \cite{Dev} :
\beqa
f_{D_8^{(1)}|_{b}}=\frac{\mbar^2}{8}\frac{\sin(\frac{\pi}{14})}
{\sin(\frac{\pi B}{14})\sin(\frac{\pi(1-B)}{14})}\label{fd81}
\eeqa
where $B$ is defined in section 2 and the mass of the particles are
related with the mass parameter $\mbar$ by : $m_8^2=m_7^2=2\mbar^2$,
 $m_a^2=8\mbar^2\sin^2(\frac{\pi a}{14})$ for $a=1,...,6$. For imaginary
coupling, one expects the particle-breather identification :
\beqa
8\mbar^2\sin^2(\frac{\pi}{14})=\big(2M\sin(\frac{\pi\xi}{14})\big)^2
\eeqa
where $\xi$ is defined as in eq. (\ref{rel}). This yields :
\beqa
f_{D_8^{(1)}|_{b=i\beta}}=-\frac{M^2}{16}\frac{\sin(\frac{\pi\xi}{14})}
{\sin(\frac{\pi}{14})\sin(\frac{\pi(1+\xi)}{14})},
\eeqa
which for $\xi=7$ is in perfect agreement with eq. (\ref{fbulk})
evaluated at $\xi=3/5$.

It is also interesting to study the behaviour of  the model (\ref{action})
for
 $p'=p+1$ in the
limit $p\rightarrow\infty$. If we consider the sine-Gordon
model with action (\ref{SG}) there is an equivalent description in terms of
the massive Thirring
model \cite{Col} :
\beqa
{\cal A}_{MT} = \int d^2x \big[{i\overline\psi}\gamma^\nu\partial_\nu\psi -
M_{SG}{\overline \psi}\psi
-\frac{g}{2}({\overline\psi}\gamma^\nu\psi)^2\big]
\eeqa
where $\psi,{\overline\psi}$ is a Dirac field. The four fermion
coupling constant $g$ relates to $\hbeta$ in (\ref{SG}) by
$\frac{g}{\pi}=\frac{1}{2\hbeta^2}-1$.
 Also the free fermion point is
reached for $\hbeta^2\rightarrow\frac{1}{2}$. Using the results of ref.
 \cite{4} concerning the one-point function $<e^{ia\phi}>$ in the SG model
one gets :
\beqa
<\cos(\hbeta\phi)>\rightarrow  \limi{\epsilon\rightarrow
0}\ M_{SG}\Gamma(\epsilon)\ \ \ \mbox{and}\ \
 \ \Lambda\rightarrow-\frac{M_{SG}}{\pi}\ \ \mbox{for}\ \ \ \
\hbeta^2\rightarrow \frac{1}{2}.\label{singul}
\eeqa
Using the boson-fermion correspondence, one has the identification
${\overline\psi}\psi\equiv \cos(\hbeta\phi)/\pi$ and
$<{\overline\psi}\psi>\rightarrow \limi{\epsilon\rightarrow
0}\ \frac{M_{SG}}{\pi}\Gamma(\epsilon)$. Let us now return to the case
 $p'=p+1$ and the limit $p\rightarrow\infty$
in (\ref{action}). It corresponds to $\beta^2\rightarrow
\frac{1}{2}$, i.e. $\xi\rightarrow 1$ and similarly to eq.
(\ref{singul}) the VEV of the perturbing
operator $\Phi^{(1)}_{12}\Phi^{(2)}_{12}$ also becomes singular.
 Consequently, the perturbing operator in (\ref{action}) can be
 conveniently rewritten in terms of two Dirac fields (with the same mass
$M_f$) of two independent massive-Thirring models at the
free fermion point :
\beqa
\lambda \Phi^{(1)}_{12}\Phi^{(2)}_{12} \rightarrow -M_f{\overline\psi}\psi
-M_f{\overline\psi}'\psi'.
\eeqa
In fact, this situation is not really surprising : when $p\rightarrow
 \infty$ the coset algebra of each ${\cal M}_{p/p+1}$ reduces to
a level-1 $SU(2)$ current algebra. The two models are coupled  via their
 primary fields in the spin $\frac{1}{2}$ representation of dimension
 $\frac{1}{4}$ \cite{muss}. As expected, in this limit the model
becomes the free field theory of two complex massive fermions with
undeformed $SO(4)$ symmetry.

Finally, let us consider the case $p'=p+1$ and $p=2$ in the action
(\ref{action}). Each minimal model is then identified to ${\cal
M}_{2/3}$ with central charge $c=0$. For each model, the unitary
representation is the vacuum one with conformal weight
$\Delta_{11}=\Delta_{12}=0$, for which cases we have the identifications
$\Phi^{(i)}_{12}={\mathbb I}^{(i)}$ for $i\in\{1,2\}$. This corresponds to
the choice
$\beta^2=\frac{1}{3}$, i.e. $\xi=\frac{1}{2}$ and, using eqs.
 (\ref{VEV2}), (\ref{M}), (\ref{fbulk}), one can check that
\beqa
<0|\Phi^{(1)}_{12}\Phi^{(2)}_{12}|0>=1\ \ \ \ \ \ \ \mbox{and}\ \ \ \ \ \
 f_{12}=\lambda,
\eeqa
as expected.\\

Let us now turn to the model associated with action (\ref{actiontilde}).
Due to eq. (\ref{restrict}) it corresponds to the choice $\beta^2=\beta^2_{-}$. 
The condition $4p>3p'$ guarantees that the perturbing operator is relevant.
Then, the vacuum structure is expected to be similar to that of (\ref{action}).
From eqs. (\ref{d}) and (\ref{VEV1}) and the substitutions :
\beqa
p\leftrightarrow p', \ \ \ \ \ (r,r')\leftrightarrow (s,s'),\ \ \ \ \ \xi
\longrightarrow
 \frac{1+\xi}{3\xi-1}\label{a'}
\eeqa
the result for the VEV immediatly follows :
\beqa
<0_\vac|\Phi^{(1)}_{rs}(x)\Phi^{(2)}_{r's'}(x)|0_\vac> &=&
{\tilde d}_{rs,r's'}^{\vac}
\Big[\frac{-\pi{\hat\lambda}\gamma(\frac{3\xi-1}{4\xi})(2\xi)^{\frac{5\xi-3}{2\xi}}}
{\gamma(\frac{1}{\xi})\gamma(\frac{\xi-1}{2\xi})}\Big]^{\frac{4\xi}{5\xi-3}(\Delta_{rs}+\Delta_{r's'})}
 \nonumber
\\
&&\times \exp{\cal Q}_{21}((1+\xi)r-2\xi s,(1+\xi)r'-2\xi s').\label{VEV1tilde}
\eeqa
The function  \ ${\cal Q}_{21}(\th,\th')$ \ is given by the integral :
\beqa
{\cal Q}_{21}(\th,\th')= \int_0^{\infty} \frac{dt}{t}
\Big(\frac{\Psi_{21}(\th,\th',t)}{\sinh((1-\xi)t)\sinh(2t\xi)\sinh((3-5\xi)t)}
-\frac{\th^2+{\th'}^2 - 2(1-\xi)^2}{4\xi(\xi+1)}e^{-2t}
\Big)\nonumber
\eeqa
with
\beqa
\Psi_{21}(\th,\th',t)&=&\Big[\cosh((\th+\th')t)\cosh((\th-\th')t) -
\cosh((2-2\xi)t)\Big]\nonumber \\
&&\ \ \ \ \times\sinh((1-\xi)t)\cosh((3-5\xi)t)\nonumber\\
&& - \ \Big[\cosh((\th+\th')t)+\cosh((\th-\th')t)-\cosh((2-2\xi)t)-1\Big]\nonumber\\
&&\ \ \ \times \ 
\sinh((3\xi-1)t/2)\cosh((1+\xi)t/2)\nonumber
\eeqa
and defined by analytic continuation outside this domain.
The prefactor associated with the QG restriction 
${\tilde d}_{rs,r's'}^{j{\tilde j}}=d_{sr,s'r'}^{j{\tilde
j}}|_{p \leftrightarrow p'}$. For $(r,s)=(r',s')=(2,1)$ (\ref{VEV1tilde})
becomes :
\beqa
<0_{\vac}|\Phi^{(1)}_{21}(x)\Phi^{(2)}_{21}(x)|0_{\vac}> &=&
\frac{1}{\hat\lambda}
\Big[\frac{-\pi{\hat\lambda}\gamma(\frac{3\xi-1}{4\xi})}
{\gamma(\frac{1}{\xi})\gamma(\frac{\xi-1}{2\xi})}
\Big]^{\frac{4\xi}{5\xi-3}}
\frac{2^{\frac{5-3\xi}{5\xi-3}}(4\xi)}{\pi(3-5\xi)}
\frac{\gamma(\frac{1+\xi}{10\xi-6})\gamma(\frac{3\xi-1}{10\xi-6})}
{\gamma(\frac{2\xi}{5\xi-3})}\label{VEV2tilde}
\eeqa
with the relation between the mass of the lightest kink $M$ and $\hat\lambda$ :
\beqa
M=\frac{2^{\frac{1+\xi}{5\xi-3}} \Gamma(\frac{1+\xi}{10\xi-6})
\Gamma(\frac{3\xi-1}{10\xi-6})}{\pi\Gamma(\frac{2\xi}{5\xi-3})}\Big[
\frac{-\pi{\hat\lambda}\gamma(\frac{3\xi-1}{4\xi})
}{\gamma(\frac{1}{\xi})\gamma(\frac{\xi-1}{2\xi})}
\Big]^{\frac{2\xi}{5\xi-3}}.\label{Mtilde}
\eeqa
For the coupled minimal models defined by (\ref{actiontilde}), the
massive phase corresponds to the domain :
\beqa
(iii) \ \ \ \frac{3}{5}\ < \xi <1, \ \ \ \ \lambda<0\ \ \ \mbox{i.e.} \ \ \
\frac{3}{4}\ <\  \frac{p}{p'}\ <\  1.
\eeqa
One also obtains the bulk free energy associated with action
(\ref{actiontilde}) :
\beqa
f_{21}=-\frac{M^2\sin(\frac{\pi(3\xi-1)}{10\xi-6})}{2}
\frac{\sin(\frac{\pi(1+\xi)}{10\xi-6})}{\sin(\frac{\pi(2\xi)}{5\xi-3})}\label{fbulktilde}.
\eeqa

{\small\section{Application and concluding remarks}}
Accepting the conjectures (\ref{VEV1}) and (\ref{VEV1tilde}), one can 
easily deduce interesting predictions\,\footnote{The same method was
applied in \cite{4} to obtain a prediction about the long distance
asymtotic of two-point correlation function
$<\sigma_0\sigma_n>_{n\rightarrow\infty}$ in the XXZ spin chain.}
 for the short and long distance asymptotic of two-point
correlation functions in the model (\ref{action}) or  (\ref{actiontilde}). For each case
depicted in figure 1, we can  express the result in terms of :
\beqa
<0_\vac|\Phi^{(1)}_{rs}(x)\Phi^{(2)}_{r's'}(x)|0_\vac>={\cal
G}_{rs,r's'}^{j{\tilde j}}
\eeqa
given by eq. (\ref{VEV1}) or (\ref{VEV1tilde}), respectively. First,
 using the short distance approximation
 (operator product expansion) and eq.(\ref{prop})
 we get \ $e^{ia\varphi_1(x)} e^{ib\varphi_2(y)} \sim
e^{i(a\varphi_1(x)+b\varphi_2(y))}$. Then, using the Coulomb gas
representation of each primary operator (\ref{primaire}) one has
\beqa
<0_\vac|\Phi^{(1)}_{rs}(x)\Phi^{(2)}_{r's'}(y)|0_\vac>\ \
\fleche{|x-y|\rightarrow 0}\ \
<0_\vac|\Phi^{(1)}_{rs}(x)\Phi^{(2)}_{r's'}(x)|0_\vac>.
\eeqa
In this limit - case $(a)$ - , the two-point functions
 then become :
\beqa
<0_\vac|\Phi^{(1)}_{rs}(x)\Phi^{(2)}_{r's'}(y)|0_\vac>\
\fleche{|x-y|\rightarrow 0}\ {\cal
G}_{rs,r's'}^\vac.\ \
\eeqa
Secondly, in the long distance approximation the asymptotic two-point
 function simply reduces to the product of two one-point functions as
$<e^{i\vaa.\vph(x)}e^{i\vab.\vph(y)}>\rightarrow <e^{i\vaa.\vph(x)}>
<e^{i\vab.\vph(y)}>$ \ when
$|x-y|\rightarrow \infty$. Then we obtain - case $(b)$ and $(d)$ - 
\beqa
&&<0_\vac|\Phi^{(1)}_{rs}(x)\Phi^{(2)}_{r's'}(y)|0_\vac>\ \
\fleche{|x-y|\rightarrow \infty}\ \
<0_\vac|\Phi^{(1)}_{rs}(x){\mathbb I}^{(2)}|0_\vac>
<0_\vac|{\mathbb I}^{(1)}\Phi^{(j)}_{r's'}(x)|0_\vac>;\nonumber\\
&&<0_\vac|\Phi^{(1)}_{rs}(x)\Phi^{(1)}_{r's'}(y)|0_\vac>\ \
\fleche{|x-y|\rightarrow \infty}\ \
<0_\vac|\Phi^{(1)}_{rs}(x){\mathbb I}^{(2)}|0_\vac>
<0_\vac|\Phi^{(1)}_{r's'}(x){\mathbb I}^{(2)}|0_\vac>
.\nonumber
\eeqa
Indeed, it gives :
\beqa
<0_\vac|\Phi^{(1)}_{rs}(x)\Phi^{(i)}_{r's'}(y)|0_\vac>\
\fleche{|x-y|\rightarrow
\infty} \
 {\cal G}_{rs,11}^\vac\big[\delta_{i1}{\cal G}_{r's',11}^\vac
+\delta_{i2}{\cal G}_{11,r's'}^\vac\big] \ \ \ \mbox{for} \ \
i\in\{1,2\}
\eeqa
with $\delta_{ii'}$ the Kr\"onecker symbol. Obsviously similar results are
obtained for the two-point function
$<0_\vac|\Phi^{(i)}_{rs}(x)\Phi^{(2)}_{r's'}(y)|0_\vac>$ using the
 ${\mathbb Z}_2$ symmetry ($1\leftrightarrow 2$).

Finally, using the fusion rules in the short distance approximation (the
two primary fields belong to the same space of states) \
$\Phi^{(i)}_{rs}\times
\Phi^{(i)}_{r's'}$ $\rightarrow$
 $\Phi^{(i)}_{r''s''}$, the two-point function is expanded in terms of
 the one-point functions
$<0_\vac|\Phi^{(1)}_{r''s''}(x){\mathbb I}^{(2)}|0_\vac>$\ or \
$<0_\vac|{\mathbb
I}^{(1)}\Phi^{(2)}_{r''s''}(x)|0_\vac>$ for $i\in\{1,2\}$ respectively.
Then, we have for $i=1$ - case $(c)$
\beqa
<0_\vac|\Phi^{(1)}_{rs}(x)\Phi^{(1)}_{r's'}(y)|0_\vac>\
\fleche{|x-y|\rightarrow
0}\
\sum_{r''s''}C_{rs,r's'}^{r''s''}
|x-y|^{2(\Delta_{r''s''}-\Delta_{rs}-\Delta_{r's'})}
{\cal G}_{r''s'',11}^{\vac}
\eeqa
and similarly for $i=2$ \ where \ $C_{rs,r's'}^{r''s''}$ \ are the structure
constants of  the minimal model operator algebra \cite{Fateev}.\\
\\
\vspace{0.3cm}
\centerline{\bf Example 1 : Two magnetically coupled Ising models}
It is now straightforward to compute different two-point correlation
functions in
one of the simplest (non-trivial) cases : two-magnetically coupled Ising
models
in the massive phase. It corresponds to $\beta^2=3/8$ i.e.
$\xi=3/5$ in (\ref{VEV1}). In this case we have the identification
$\Phi^{(i)}_{12}=\sigma^{(i)}$ with conformal dimension
$\Delta_\sigma=1/16$ - the spin operator - and
$\Phi^{(i)}_{13}=\epsilon^{(i)}$
with conformal dimension
$\Delta_\epsilon=1/2$ - the energy operator. There are two degenerate ground
states \cite{vays,muss} denoted $|0_{00}>\equiv|->$ and  $|0_{1/2\
1/2}>\equiv|+>$.
For simplicity, we write $<\pm|...|\pm>\equiv<...>_\pm$. Sometimes, the
reflection relation :
\beqa
{\cal G}(\eta_1,\eta_2)=S_L(\eta_2){\cal G}(\eta_1,-\eta_2+2\beta-1/\beta)
\eeqa
with
\beqa
S_L(\eta_2)=\big[
-\pi\mu'\gamma(-2\beta^2)\big]^{\frac{-\eta_2\beta+\beta^2-1/2}{\beta^2}}\times
\frac{\Gamma(-2\eta_2\beta+2\beta^2) \Gamma(\eta_2/\beta+1/2\beta^2)}
{\Gamma(2+2\eta_2\beta-2\beta^2) \Gamma(2-\eta_2/\beta-1/2\beta^2)}\nonumber
\eeqa
is useful for analytic continuation. Using eq. (\ref{VEV1}) and
$d_{11,1s}^\pm=d_{1s,11}^{\pm}$
for all $s$ we obtain for instance :
\beqa
<\sigma^{(i)}>_{\pm}&=&{\cal G}_{11,12}^\pm  = \pm
1.297197220...(-\lambda)^{1/14};\nonumber\\
<\epsilon^{(i)}>_{\pm}&=&{\cal G}_{11,13}^\pm  = -
2.278284275...(-\lambda)^{4/7};\nonumber\\
<\sigma^{(1)}(0)\sigma^{(2)}(0)>_{\pm}&=&{\cal G}_{12,12}^\pm  =
1.698928047...(-\lambda)^{1/7};\nonumber\\
<\sigma^{(1)}(0)\sigma^{(2)}(\infty)>_{\pm}&=& ({\cal G}_{11,12}^\pm)^2 =
1.682720628...(-\lambda)^{1/7};\nonumber\\
<\sigma^{(1)}(0)\epsilon^{(2)}(0)>_{\pm}&=&{\cal G}_{12,13}^\pm  = \mp
3.311880669...(-\lambda)^{9/14};\nonumber\\
<\sigma^{(1)}(0)\epsilon^{(2)}(\infty)>_{\pm}&=&{\cal G}_{11,12}^\pm{\cal
G}_{11,13}^\pm
 =\mp 2.955384028...(-\lambda)^{9/14};\nonumber\\
<\epsilon^{(1)}(0)\epsilon^{(2)}(\infty)>_{\pm}&=&({\cal G}_{11,13}^\pm)^2
 = 5.160349412...(-\lambda)^{8/7};\nonumber
\eeqa
where the parameter $\lambda$ is related to the mass of the lowest kink
by :
\beqa
\lambda=-0.2379062104...M^{7/4}.
\eeqa
Notice that $<\sigma^{(1)}(0)\sigma^{(2)}(0)>_{\pm}$ and
$<\sigma^{(1)}(0)\sigma^{(2)}(\infty)>_{\pm}$ differ by less than
$0.7\%$ as expected\,\footnote{For instance, it is known that form
factors are able to reproduce with high accuracy the UV behaviour of the
correlation functions \cite{3,yu,form}.}.

\vspace{0.3cm}

\centerline{\bf Example 2 : Two energy-energy coupled tricritical Ising
models}
\vspace{0.3cm}
The case $p=4$, $p'=5$ in (\ref{action}) describes two tricritical Ising
models which interact through their leading energy density operators
$\Phi_{12}^{(i)}=\epsilon^{(i)}$ of conformal dimension
$\Delta_{\epsilon}=1/10$.  It corresponds to \ $\beta^2=2/5$\ i.e.\
\ $\xi=2/3$\  in (\ref{VEV1}). Beside $\epsilon^{(i)}$ and the identity operator,
each minimal model contains the sub-leading energy density operator
$\Phi_{13}^{(i)}={\epsilon'}^{(i)}$ with $\Delta_{\epsilon'}=3/5$ (``vacancy
operator'') , two magnetic operators $\Phi_{22}^{(i)}=\sigma^{(i)}$ with
$\Delta_{\sigma}=3/80$, $\Phi_{21}^{(i)}={\sigma'}^{(i)}$ with
$\Delta_{\sigma'}=7/16$ and $\Phi_{14}^{(i)}$. Due to the obvious property
$d_{rs,r's'}^{\vac}=d_{11,r's'}^{\vac}d_{rs,11}^{\vac}$, similarly to the
previous case we find for instance for any vacuum $|\vac>$ :
\beqa
<\sigma^{(1)}>_{\vac}&=&{\cal G}_{22,11}^\vac  = d_{22,11}^\vac\times
1.144656674...(-\lambda)^{3/64};\nonumber\\
<\epsilon^{(1)}>_{\vac}&=&{\cal G}_{12,11}^\vac  = d_{12,11}^\vac\times
1.529866659...(-\lambda)^{1/8};\nonumber\\
<\sigma^{(1)}(0)\sigma^{(2)}(0)>_{\vac}&=&{\cal G}_{22,22}^\vac  =
d_{22,22}^\vac \times 1.315726811...(-\lambda)^{3/32};\nonumber\\
<\sigma^{(1)}(0)\sigma^{(2)}(\infty)>_{\vac}&=& {\cal
G}_{22,11}^\vac{\cal G}_{11,22}^\vac = d_{22,22}^\vac \times
 1.310238901...(-\lambda)^{3/32};\nonumber\\
<\epsilon^{(1)}(0)\epsilon^{(2)}(0)>_{\vac}&=&{\cal G}_{12,12}^\vac  =
d_{12,12}^\vac \times 2.419476973...(-\lambda)^{1/4};\nonumber\\
<\epsilon^{(1)}(0)\epsilon^{(2)}(\infty)>_{\vac}&=& {\cal
G}_{12,11}^\vac{\cal G}_{11,12}^\vac = d_{12,12}^\vac \times
2.340491994...(-\lambda)^{1/4};\nonumber
\eeqa
where the parameter $\lambda$ is related to the mass of the lowest kink
by :
\beqa
\lambda=-0.2566343706...M^{8/5}.
\eeqa
Notice that $<\sigma^{(1)}(0)\sigma^{(2)}(0)>_{\vac}$ and
$<\sigma^{(1)}(0)\sigma^{(2)}(\infty)>_{\vac}$,\
$<\epsilon^{(1)}(0)\epsilon^{(2)}(0)>_{\vac}$ and
 $<\epsilon^{(1)}(0)\epsilon^{(2)}(\infty)>_{\vac}$ differ by less than
 $0.5\%$ and $3\%$ respectively.\\
\vspace{0.3cm}

\centerline{\bf Example 3 : Two coupled $A_5$ RSOS models}
\vspace{0.2cm}
The case $p=5$, $p'=6$ in (\ref{action}) describes two $A_5$ RSOS
models coupled by their primary operators
$\Phi_{12}^{(i)}$ with conformal dimension
$\Delta_{12}=1/8$.  It corresponds to $\beta^2=5/12$ i.e.
$\xi=5/7$  in (\ref{VEV1}). Each minimal model also
contains the primary operator $\Phi_{22}^{(i)}$ with
$\Delta_{22}=1/40$. As before we obtain :
\beqa
<\Phi_{22}^{(1)}>_{\vac}&=&{\cal G}_{22,11}^\vac  = d_{22,11}^\vac\times
1.090446894...(-\lambda)^{1/30};\nonumber\\
<\Phi_{12}^{(1)}>_{\vac}&=&{\cal G}_{12,11}^\vac  = d_{12,11}^\vac \times
1.726352342...(-\lambda)^{1/6};\nonumber\\
<\Phi_{22}^{(1)}(0)\Phi_{22}^{(2)}(0)>_{\vac}&=&{\cal G}_{22,22}^\vac  =
d_{22,22}^\vac \times 1.191588988...(-\lambda)^{1/15};\nonumber\\
<\Phi_{22}^{(1)}(0)\Phi_{22}^{(2)}(\infty)>_{\vac}&=& {\cal
G}_{22,11}^\vac{\cal G}_{11,22}^\vac = d_{22,22}^\vac \times
1.189074429...(-\lambda)^{1/15};\nonumber
\eeqa
where the parameter $\lambda$ is related to the mass of the lowest kink
by :
\beqa
\lambda=-0.2697511940...M^{3/2}.
\eeqa
Notice that $<\Phi_{22}^{(1)}(0)\Phi_{22}^{(2)}(0)>_{\vac}$ and
$<\Phi_{22}^{(1)}(0)\Phi_{22}^{(2)}(\infty)>_{\vac}$ differ by \
 $\sim 2\%$.\\
\vspace{0.3cm}

\centerline{\bf Example 4 : Two energy-energy coupled 3-state Potts models}
\vspace{0.2cm}
The case $p=5$, $p'=6$ in (\ref{actiontilde}) describes two 3-state Potts
models coupled \cite{zN} by their energy density operator
$\Phi_{21}^{(i)}=\epsilon^{(i)}$ with conformal dimension
$\Delta_{21}=2/5$.  It corresponds to \ 
$\xi=5/7$\ in (\ref{VEV1tilde}). Each minimal model also contains the primary operator
$\Phi_{23}^{(i)}=\sigma^{(i)}$ - the spin operator - with
$\Delta_{23}=1/15$. We obtain for instance :
\beqa
<\sigma^{(1)}>_{\vac}&=&{\cal G}_{23,11}^\vac  = {\tilde d}_{23,11}^\vac\times
1.9079...(-{\hat\lambda})^{1/3};\nonumber\\
<\sigma^{(1)}(0)\sigma^{(2)}(0)>_{\vac}&=&{\cal G}_{23,23}^\vac  =
{\tilde d}_{23,23}^\vac \times 4.50...(-{\hat\lambda})^{2/3};\nonumber\\
<\sigma^{(1)}(0)\sigma^{(2)}(\infty)>_{\vac}&=& {\cal
G}_{23,11}^\vac{\cal G}_{11,23}^\vac = {\tilde d}_{23,23}^\vac \times
3.64...(-{\hat\lambda})^{2/3};\nonumber
\eeqa
where the parameter $\hat\lambda$ is related to the mass of the lowest kink
by :
\beqa
{\hat\lambda}=-0.2612863655...M^{2/5}.
\eeqa

To conclude, we would like to mention that here we studied only a
special case of a much more general class of integrable coupled models
\cite{zN,vays,muss}. There exist many other examples which can be worked out
along the same lines. Let us also note that the exact form factor
techniques as well as the truncated conformal space approach may be used and
similar numerical analyses can be performed for the correlation functions.

\vspace{1cm}

\paragraph*{Aknowledgements}
I am grateful to C. Ahn, G. Delfino, G. Delius, D. Reynaud, P. Simon, C.
Kim, C. Rim, L. Zhao and especially to V.A.
Fateev and E. Corrigan for
useful discussions. Work supported by the EU under contract ERBFMRX
CT960012 and Marie Curie fellowship HPMF-CT-1999-00094.

\end{document}